\documentclass{article}
\usepackage[utf8]{inputenc}
\usepackage{spconf,amsmath,graphicx,hyperref}
\usepackage{enumitem}
\usepackage{amsmath,amssymb,amsfonts}
\usepackage{graphicx}
\usepackage{textcomp}
\usepackage{xcolor}
\usepackage{graphicx}
\usepackage{bm}
\usepackage{multirow}
\usepackage{caption}
\usepackage{algorithm}
\usepackage{subfigure}
\usepackage{diagbox}
\usepackage{amssymb}
\usepackage{booktabs}
\usepackage{makecell}
\usepackage{adjustbox}
\usepackage{marvosym}
\usepackage{algorithmicx}
\usepackage{algpseudocode}
\usepackage{cite}
\usepackage{braket}
\usepackage{color}
\usepackage{ifsym}
\usepackage{threeparttable}

\title{Improving Anomalous Sound Detection with Attribute-aware Representation from Domain-adaptive Pre-training}

\name{\parbox{\linewidth}{\centering
Xin Fang$^{1,2*}$, Guirui Zhong$^{1*}$, Qing Wang$^{1\dagger}$, Fan Chu$^{3}$, Lei Wang$^{2}$, Mengui Qian$^{3}$, \\
Mingqi Cai$^{2}$, Jiangzhao Wu$^{3}$, Jianqing Gao$^{2}$, and Jun Du$^{1}$
}
\thanks{
  \begin{tabular}{@{}l@{}}
    $^*$Equal contribution \\
    $^\dagger$Corresponding author
  \end{tabular}
    }
}

\address{$^1$ University of Science and Technology of China, Hefei, China \\
$^2$ iFLYTEK Research, Hefei, China \\
$^3$ National Intelligent Voice Innovation Center, Hefei, China }

\begin{document}
\ninept
\maketitle
\begin{abstract}
Anomalous Sound Detection (ASD) is often formulated as a machine attribute classification task, a strategy necessitated by the common scenario where only normal data is available for training. However, the exhaustive collection of machine attribute labels is laborious and impractical. To address the challenge of missing attribute labels, this paper proposes an agglomerative hierarchical clustering method for the assignment of pseudo-attribute labels using representations derived from a domain-adaptive pre-trained model, which are expected to capture machine attribute characteristics. We then apply model adaptation to this pre-trained model through supervised fine-tuning for machine attribute classification, resulting in a new state-of-the-art performance. Evaluation on the Detection and Classification of Acoustic Scenes and Events (DCASE) 2025 Challenge dataset demonstrates that our proposed approach yields significant performance gains, ultimately outperforming our previous top-ranking system in the challenge. 
\end{abstract}
\begin{keywords}
Anomalous sound detection, domain-adaptive pre-training, pseudo-labeling, attribute-aware representation
\end{keywords}
\section{Introduction}
\label{sec:intro}

Anomalous Sound Detection (ASD) aims to identify anomalous sounds emitted from machines, which has gained extensive research interest from both academia and industry due to its substantial benefits for machine maintenance and safety protection. Given the scarcity of anomalous sounds in real-world scenarios, ASD systems are typically developed using only normal sounds, making it a Self-Supervised Learning (SSL) task\cite{wilkinghoff2024self, wang2024representation, jiang25c_interspeech,10329432}. During inference, the anomaly score of a given sample is measured by its deviation from the learned distribution of normal data.

Recently, methods based on pre-trained models have achieved state-of-the-art (SOTA) performance and become the dominant approaches in ASD \cite{jiang25c_interspeech,jiang2024anopatch,LvAITHU2024, WangMYPS2025,10889514,zhong2025enhanced}. These methods firstly pre-train a model on large-scale audio datasets, such as AudioSet, to learn universal audio representations. Then, to adapt the pre-trained model to the ASD domain, machine attribute classification is employed as a downstream task. This strategy is preferred over simple binary classification due to the absence of anomalous training data. Specifically, machine attributes correspond to various operational conditions (e.g., speed, voltage) that describe the machine's normal working states in detail. By training the model to classify unique combinations of machine types and attributes, it learns fine-grained representations of normal machine sounds. Consequently, the distance of a test sample from normal data distribution in the representation space serves as its anomaly score. While leveraging attribute information has proven effective for enhancing ASD performance, the exhaustive collection of attribute labels for machines is laborious and often impractical. 

Reflecting this challenge, the Detection and Classification of Acoustic Scenes and Events (DCASE) Challenge task 2 has focused on ASD in attribute-unavailable scenarios since 2024\cite{Nishida_arXiv2025_01}. Researchers have explored various pseudo labeling methods for attribute-unavailable machines, typically by applying clustering algorithms to different audio embeddings. For instance, one common strategy involves extracting general audio embeddings using a powerful pre-trained model\cite{10890020}. Another approach is to fine-tune pre-trained models on a machine attribute classification task, where all sounds from a machine lacking attribute information are grouped into a single class, to extract domain-specific embeddings\cite{10890020, LvAITHU2024}. However, both strategies present significant drawbacks. First, due to the domain mismatch between general audio datasets (e.g., speech and music are predominant in AudioSet\cite{gemmeke2017audio}) and industrial machine sounds, pre-trained embeddings often fail to model the fine-grained nature of machine audio. Second, by treating all sounds from an attribute-unavailable machine as one class, the fine-tuning process suppresses important intra-class variations that correspond to distinct operational attributes. This collapse of features negatively impacts the performance of subsequent clustering process.

To address these challenges, we propose a novel anomalous sound detection approach that leverages agglomerative hierarchical clustering of the attribute-aware representations derived from domain-adaptive pre-training. Based on a universal pre-training model on AudioSet, we introduce domain-adaptive pre-training using machine sound datasets, which serves two key purposes. First, it allows the model to leverage its general audio modeling capabilities while adapting to capture the essential characteristics of machine sounds more effectively, thereby providing a more fine-grained representation for subsequent clustering and fine-tuning. Second, by employing a SSL paradigm instead of classifying all data from an attribute-unavailable machine as a unique class, our method preserves crucial intra-class differences, benefiting for the clustering process of unattributed data. We highlight three key contributions of this paper as follows:
\begin{enumerate}[label=(\arabic*), itemsep=-4pt]
    \item We introduce a domain-adaptive pre-training method that leverages multiple machine sound datasets to learn fine-grained, attribute-aware representations.
    \item We propose a pseudo-labeling approach for attribute-unavailable machines, which uses agglomerative hierarchical clustering on embeddings extracted from domain-adaptive pre-training.
    \item We fine-tune the domain-adaptive pre-trained model on the downstream ASD task using the pseudo-attribute labels, achieving a new SOTA performance on the DCASE 2025 ASD evaluation dataset.
\end{enumerate}

\section{proposed method}
\label{sec:method}

\subsection{Overview}
\begin{figure}[h]
    \centering
    \includegraphics[width=1.0\linewidth]{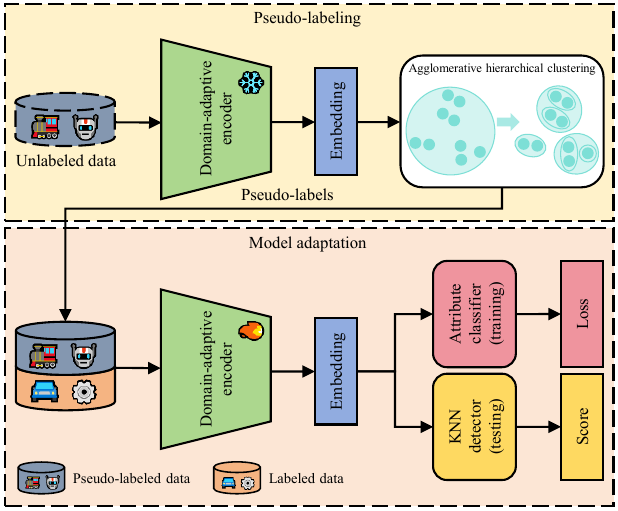}
    \vspace{-0.4cm}
    \caption{The overall framework of our proposed method.}
    \label{fig: framework}
    \vspace{-0.2cm}
\end{figure}

The proposed ASD framework, illustrated in Fig.~\ref{fig: framework}, comprises two main stages: (1) pseudo-labeling, (2) model adaptation. Specifically, the pseudo-labeling process aims to assign pseudo-attributes to attribute-unavailable machines by applying agglomerative hierarchical clustering to the embeddings produced by a domain-adaptive encoder. The domain-adaptive encoder is obtained by further pre-training a SSL model on multiple machine sound datasets. This pre-training is intended to better model the ASD domain's data and extract attribute-aware audio embeddings that preserve intra-class variations, which supports both pseudo-labeling and downstream adaptation. In the second stage, model adaptation, the domain-adaptive encoder is further fine-tuned on the downstream ASD task using both ground-truth and pseudo-attribute labels, thereby transferring the learned representations to the target task. Ultimately, the framework frames ASD as a machine attribute classification problem. The key steps are elaborated in the following subsections. 

\subsection{Review of machine attribute classification}
\label{framework}
As mentioned above, due to the characteristic of ASD, we formulate the problem as a Machine Attribute Classification (MAC) subtask, in which each distinct combination of machine type and attribute is treated as a separate class\cite{zhong2025enhanced,10888266, 10448126}. For example, ToyCar and BandSealer machines have attributes defined by speed (denoted spd); hence (ToyCar, spd28V), (ToyCar, spd31V), and (BandSealer, spd6) are regarded as three different classes.

Adapting pre-trained models to the ASD domain through MAC has shown great advantages\cite{LvAITHU2024, WangMYPS2025, SaengthongSCITOK2025}. These methods typically use a Vision Transformer (ViT) backbone pre-trained on AudioSet and fine-tune it on machine audio through MAC to transfer general audio representations to the machine audio domain. The fine-tuned encoder maps machine audio into a representation space, and an anomaly detector based on K-Nearest Neighbors (KNN) computes anomaly scores from the embedding distance between a test example and its nearest normal example in the training set. Given an audio spectrogram $\mathbf{X}\in\mathbb{R}^{{F}\times{T}}$, where $F$ denotes the frequency dimension and $T$ denotes the time dimension, the fine-tuning stage can be defined below:
\begin{gather}
\label{eq: embedding}
    \mathbf{E} = \mathcal{F}(\mathbf{X}) \\
\label{eq: arcface}
    \hat{l}_\mathrm{attr} = \mathcal{C}_\mathrm{attr}(\mathbf{E}) \\
\label{eq: loss}
    \mathcal{L}_\mathrm{ASD} = \mathrm{CE}(\hat{l}_\mathrm{attr}, {l}_\mathrm{attr})
\end{gather}
where $\mathbf{E}$ and $\mathcal{F}$ denote the embedding of machine audio and the encoder function, separately. $\mathcal{C}_\mathrm{attr}$ indicates the ArcFace attribute classifier\cite{deng2019arcface}, $\mathrm{CE}$ is the cross-entropy loss, and $\mathcal{L}_\mathrm{ASD}$ is the ASD loss to be optimized. ${l}_\mathrm{attr}$ and $\hat{l}_\mathrm{attr}$ represent the ground-truth and predicted attribute labels. After fine-tuning, the embedding will pass through KNN backend to obtain anomaly score for evaluation.

\subsection{Domain-adaptive pre-training}
\label{dap}
Domain-adaptive pre-training serves as a critical step, providing attribute-aware representations that facilitate subsequent unsupervised clustering and downstream fine-tuning. The optimization in our domain-adaptive pre-training follows that of the Efficient Audio Transformer (EAT)\cite{chen2024eat}; the principal differences are the datasets and the downstream objectives. We briefly introduce the pre-training process of EAT. The full details can be found in the original paper\cite{chen2024eat}.

EAT uses a student-teacher architecture in which both branches are ViT encoders. Inputs are patch-level spectrograms, and the student branch receives a proportionally masked spectrogram. EAT introduces an Utterance-Frame Objective (UFO) loss function and updates the teacher parameters via Exponential Moving Average (EMA) of the student parameters. For the frame-level loss, the student branch performs the Masked Auto-Encoder\cite{9879206} reconstruction, producing decoder outputs $\mathbf{X}_\mathrm{o}$, while the teacher branch generates averaged ViT layer outputs $\mathbf{Y}_\mathrm{o}$. So the frame-level loss can be computed as:
\begin{gather}
\mathcal{L}_\mathrm{f} = \Vert\mathbf{X}_\mathrm{o} - \mathbf{Y}_\mathrm{o}\Vert_2^2
\end{gather}
Moreover, EAT incorporates a learnable classification token (CLS token) $\mathbf{c}$ into the student branch to capture global utterance-level information. Meanwhile, the teacher branch produces a corresponding global representation $\mathbf{y}$ by averaging its ViT outputs across the layer and patch dimensions. The utterance-level loss is then calculated as:
\begin{gather}
\mathcal{L}_\mathrm{u} = \Vert\mathbf{c} - \mathbf{y}\Vert_2^2
\end{gather}
Finally, the UFO loss is defined as below:
\begin{gather}
\mathcal{L}_\mathrm{UFO} = \mathcal{L}_\mathrm{f} + \mathcal{L}_\mathrm{u}
\end{gather}

Following pre-training on AudioSet, we employ the aforementioned updating strategy to further optimize the model on DCASE machine sound data. This step helps mitigate performance degradation arising from the domain mismatch between AudioSet and machine audio. Furthermore, because the SSL approach operates without labels, the model is expected to learn the fine-grained variations corresponding to distinct attributes within the same machine type. These capabilities facilitate the subsequent pseudo-labeling and model adaptation processes. 

Additionally, we scale up the machine sound data in the domain-adaptive pre-training stage by using the datasets from previous DCASE ASD challenges to further improve both clustering and ASD performance. 

\subsection{Pseudo-labeling with attribute-aware embedding}
In the MAC task, machines of the same type that lack attribute information are typically grouped into a single class, e.g., (ToyCar, noAttr). This coarse-grained classification limits further improvements in ASD system performance because it overlooks latent attribute variations within the same machine type. To address this, several works have focused on assigning pseudo-labels to machines without attributes. These approaches can be broadly divided into two categories: 1) analysis via mechanism indicators, and 2) clustering of model embeddings.

1) Analysis via mechanism indicators: Our top-ranking system in the DCASE 2025 ASD challenge employed distinct mechanism indicators for attribute analysis\cite{WangMYPS2025}. Taking the ScrewFeeder as an example, frequency domain analysis revealed prominent periodic acoustic signatures in the 1-3kHz band. Based on this analysis, we assigned attributes according to the derived periodic parameters. Although this approach can yield effective results, its reliance on carefully engineered, device-specific indicators makes it difficult and inconvenient to implement.

2) Clustering of model embeddings: These methods differ primarily in how they acquire embeddings. Some approaches use external pre-trained models to directly extract machine audio embeddings\cite{10890020}. Others, such as the top-ranking system in the DCASE 2024 Challenge\cite{LvAITHU2024}, first fine-tune a pre-trained model on the target data before extracting embeddings. However, both strategies have drawbacks. The former often suffers from poor modeling capability due to the significant domain mismatch between the pre-training data and the target machine sounds. The latter risks suppressing crucial intra-class attribute differences, as the fine-tuning stage treats all samples from an attribute-unavailable machine type as a single class. As discussed in Section \ref{dap}, our proposed domain-adaptive pre-training stage effectively alleviates both of these drawbacks.

Our domain-adaptive pre-training yields a model with a powerful modeling capability for modeling machine sounds. This model can extract fine-grained, attribute-aware embeddings that capture the intrinsic nature of the machine sound. We refer to the domain-adaptive ViT encoder as $\mathcal{F}_\mathrm{DA}$, and the resulting embedding $\mathbf{E}_\mathrm{DA}$ used for clustering is defined as:
\begin{gather}
\label{eq: FDA}
    \mathbf{Z} = \mathcal{F}_\mathrm{DA}(\mathbf{X}) \\
\label{eq: EDA}
    \mathbf{E}_\mathrm{DA} = \frac{1}{\mathrm{P}}\sum_{\mathrm{i}=1}^\mathrm{P}\mathbf{z}_\mathrm{i}
\end{gather}
where $\mathbf{Z} = 
[\mathbf{z}_1, \mathbf{z}_2, ..., \mathbf{z}_\mathrm{P}]^T\in\mathbb{R}^{{P}\times{D}}$. $P$ and $D$ denote the number of patches and the embedding dimension.

As shown in Fig.~\ref{fig: framework}, with the attribute-aware embeddings $\mathbf{E}_\mathrm{DA}$ from the domain-adaptive encoder, we apply agglomerative hierarchical clustering using Ward linkage\cite{murtagh2012algorithms}. Specifically, the Ward linkage utilizes the Error Sum of Squares (ESS) quantity as a criterion to measure the information loss when merging clusters. For a given cluster $\mathcal{C}$, ESS is defined as follows:
\begin{gather}
\label{eq: ward}
    \mathrm{ESS}(\mathcal{C}) = \sum_{\mathbf{x}\in\mathcal{C}}
    (\mathbf{x}-\mathbf{m}_{x})^T(\mathbf{x}-\mathbf{m}_{x})
\end{gather}
where $\mathbf{m}_{x}$ is the mean of the samples in cluster $\mathcal{C}$. The objective of Ward linkage is to minimize the ESS (information loss) at each merging step. This process yields compact and uniform clusters, which are then treated as the pseudo attributes.

\subsection{Model adaptation with attribute information}
In the final model adaptation stage, we fine-tune the domain-adaptive pre-trained model on the MAC task. This step uses both the newly generated pseudo attributes $\mathcal{A}_\mathrm{p}$ and the available ground-truth attributes $\mathcal{A}_\mathrm{g}$ to further adapt the model to the ASD domain. Therefore, for a given machine type $\mathcal{M}$, the corresponding attribute labels ${l}_\mathrm{attr}$ in Equation \ref{eq: loss} is defined as follows:
\begin{gather}
\label{eq: label}
    {l}_\mathrm{attr} = (\mathcal{M}, \mathcal{A})
\end{gather}
For machines lacking ground-truth attributes, their assigned attribute $\mathcal{A}\in\mathcal{A}_\mathrm{p}$. To validate our method's effectiveness for machines lacking attributes, we constructed a separate evaluation set (``NoAttr set'') composed exclusively of samples from these machines.

Although the dominant approach of first pre-training a ViT on AudioSet and then fine-tuning it on DCASE data  has achieved great performance, it suffers from a key drawback: the significant domain mismatch between general audio and machine sounds hinders further improvements. The domain-adaptive pre-training stage is designed to bridge this gap, facilitating a smoother transfer of pre-trained knowledge to the ASD domain.  Hence, we implement the fine-tuning process detailed in Section \ref{framework}, updating the domain-adaptive pre-trained model with both the pseudo-attribute groups  $\mathcal{A}_\mathrm{p}$ and ground-truth attribute groups $\mathcal{A}_\mathrm{g}$.

\section{experiments}
\label{sec:typestyle}
\subsection{Dataset and evaluation metric}
We evaluate our method on the DCASE 2025 ASD dataset\cite{Harada2021, Dohi2022}, which contains a development set (seven machine types) and an additional set (eight machine types). Seven of these machine types are provided without explicit attributes. For each machine type, the dataset includes 1,000 training clips and 200 test clips. The training data is divided into 990 source-domain clips and 10 target-domain clips. The test data comprises 100 clips for the source domain and an additional 100 clips for the target domain, in which there are 50 normal and 50 anomalous clips, respectively.  A key challenge is that machine attributes are not shared between the source and target domains. Meanwhile, we evaluate ASD performance with the harmonic mean of two metrics: the Area Under the Receiver Operating Characteristic (ROC) Curve (AUC) and the partial AUC (pAUC) with $p=0.1$. This score is calculated from the system's output anomaly scores. The harmonic mean on the evaluation set serves as the official score according to the challenge rules\cite{Nishida_arXiv2025_01}.

\subsection{Implementation details}
The overall framework used in this work is based on the public open-source EAT project\cite{chen2024eat}. All audio waveforms are padded or truncated to a length of 10s, and then converted to log-mel spectrograms with a frame length of 25ms, a frame shift of 10ms, and 128 mel bins. For the domain-adaptive pre-training stage, the hyperparameters largely follow those of the original EAT framework, with the exception of the batch size, which we set to 32. In the fine-tuning stage, the model is trained for 20 epochs with a batch size of 32. A cosine learning rate scheduler is adopted with an upper limit of 5e-5 and a warm-up step of 120. Mixup\cite{zhang2017mixup} and SpecAugment\cite{park19e_interspeech} data augmentation methods are used during training. We conducted all experiments three times independently and reported the mean and standard deviation.

\begin{table}[t]
	\renewcommand\arraystretch{1.25}
	\newcolumntype{L}[1]{>{\raggedright\arraybackslash}p{#1}}
	\newcolumntype{C}[1]{>{\centering\arraybackslash}p{#1}}
	\newcolumntype{R}[1]{>{\raggedleft\arraybackslash}p{#1}}
	\centering
        \footnotesize
	\caption{Comparison of ASD performance using pseudo-attribute labels generated from various embedding models.}
	\vspace{-0.3cm}
	\label{tab: pseudo labels}\medskip
	\resizebox{8.5 cm}{!}
        {
        \begin{threeparttable} 
        \begin{tabular}{c | c c c c}
            \toprule[1 pt]
            {Scheme} & {Dev set} & {Eval set} & {NoAttr set} & {All}\\
            \midrule
            No.1* & 59.18 
            & \textbf{61.62} 
            & \textbf{65.60} 
            & 60.46  \\
            \midrule
            N/A & 60.41{\scriptsize$\pm$0.96} 
            & 58.23{\scriptsize$\pm$0.35} 
            & 62.13{\scriptsize$\pm$1.57} 
            & 59.22{\scriptsize$\pm$0.35}  \\

            GP & 59.29{\scriptsize$\pm$0.46} 
            & 58.19{\scriptsize$\pm$0.50} 
            & 61.08{\scriptsize$\pm$0.56} 
            & 58.69{\scriptsize$\pm$0.16}  \\

            FT & 59.97{\scriptsize$\pm$0.75} 
            & 59.75{\scriptsize$\pm$0.52} 
            & 62.75{\scriptsize$\pm$0.49} 
            & 59.85{\scriptsize$\pm$0.61}  \\
            \midrule
            DAP & 61.11{\scriptsize$\pm$0.39} 
            & 60.32{\scriptsize$\pm$1.09} 
            & 64.14{\scriptsize$\pm$0.09} 
            & 60.67{\scriptsize$\pm$0.41}  \\

            DAP-full & \textbf{62.05}{\scriptsize$\pm$0.29} 
            & 60.28{\scriptsize$\pm$0.43} 
            & 65.41{\scriptsize$\pm$0.14}
            & \textbf{61.09}{\scriptsize$\pm$0.33}   \\
            \bottomrule[1 pt]
    \end{tabular}
    \begin{tablenotes}[para,flushleft]
    \footnotesize              
    \item* The challenge does not provide the mean and standard deviation values. 
  \end{tablenotes}         
\end{threeparttable}
    }
\vspace{-0.2 cm}
\end{table}

\subsection{Experimental results and analysis}
Table~\ref{tab: pseudo labels} presents an ablation study on ASD performance using pseudo-attribute labels generated from different embedding sources. To ensure a fair comparison and isolate the impact of the pseudo labels, all models in Table~\ref{tab: pseudo labels} are fine-tuned from the same baseline: a standard EAT model pre-trained on AudioSet. ``NoAttr set'' composes of samples from machines lacking attributes. We compare our proposed Domain-Adaptive Pre-training (DAP) against several baselines: without using pseudo labels (N/A), the standard EAT model pre-trained on AudioSet (GP), and its fine-tuned version (FT). The results clearly show that using pseudo labels from our DAP scheme yields the best ASD performance across all machines, outperforming the N/A, GP, and FT baselines. This finding demonstrates the superiority of pseudo-labeling using embeddings from our domain-adaptive approach. Moreover, to investigate the effect of data scaling, we introduce an enhanced model, ``DAP-full'', by performing domain-adaptive pre-training on the combined DCASE 2020–2025 ASD datasets. DAP-full further achieves the best score across all machines and comparable performance with our top-ranking challenge scheme on the NoAttr set, demonstrating the benefits brought by scaling up data in the DAP process.

\begin{figure}[htbp]
	\centering
	\begin{minipage}{0.49\linewidth}
		\centering
		\includegraphics[width=1.0\linewidth]{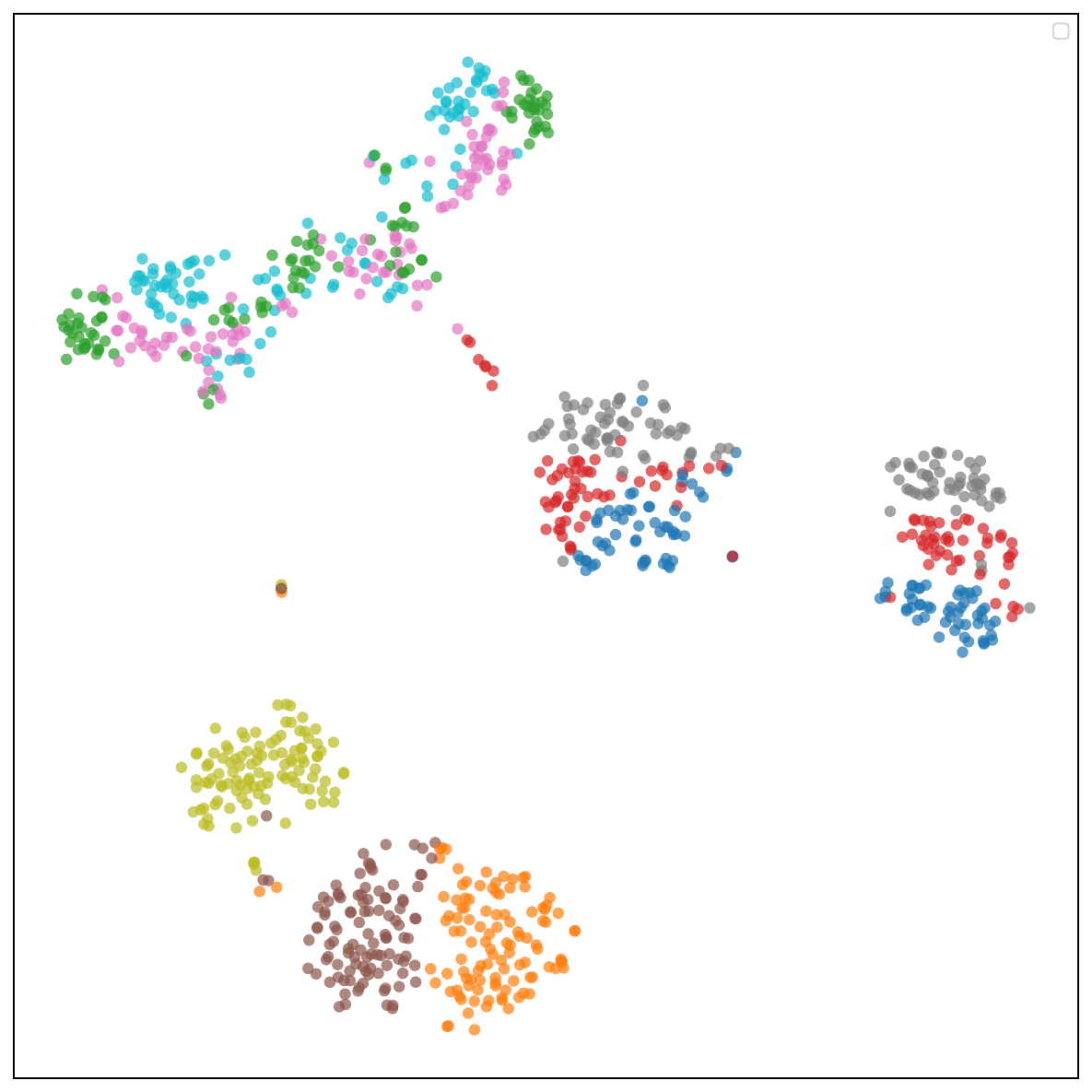}
	\end{minipage}
	\begin{minipage}{0.49\linewidth}
		\centering
		\includegraphics[width=1.0\linewidth]{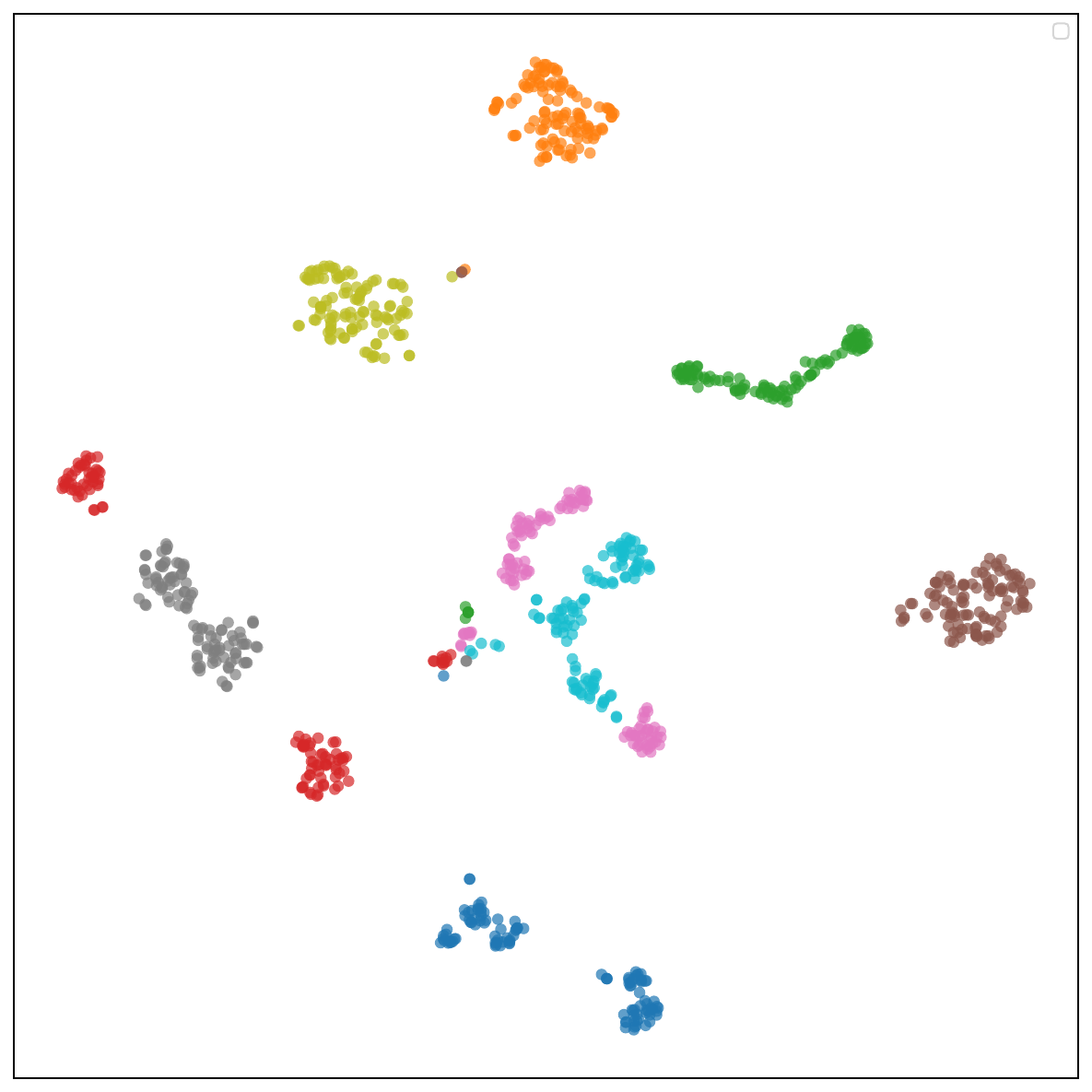}
	\end{minipage}
    
    \caption{T-SNE Visualization of the embedding distribution of FT (left) and DAP (right) schemes. Different colors represent different real attributes of machine type Polisher in the source domain, such as (pow1, nA) and (pow3, nB). The pow (power) and n (noise) denote the Polisher's working power and background noise, and the characters after pow and n represent their specific attribute values.}
    \label{visual}
\end{figure}

To validate the quality of the embeddings generated by our DAP scheme, we use t-SNE to visualize the embedding distribution for the ``Polisher'' machine type in the source domain, as shown in the Fig. \ref{visual}. The visualization reveals that embeddings from the conventional FT scheme exhibit significant overlap and unclear boundaries between different machine attributes. In contrast, our DAP scheme produces more separable clusters by enhancing both intra-class compactness and inter-class separation. This improved representation reflects the differences within the same machine type, which is crucial for the effectiveness of the subsequent pseudo-labeling process.

Further, we conduct experiments to evaluate the benefits of our Model Adaptation (MA) strategy. As shown in Table \ref{tab: ablation}, MA with the DAP-full scheme improves the overall score by 2.06\% compared to a vanilla approach (i.e., fine-tuning directly from the general pre-trained model), confirming that our domain-adaptive stage enables a more effective transfer of knowledge from general audio to machine sound. Additionally, we incorporated the generated pseudo-labels during this adaptation. As shown in the last row of Table \ref{tab: ablation}, this step provides an further performance gain of 1.05\%, leading to the best overall results.

\begin{figure}[t]
    \centering
    \includegraphics[width=0.9\linewidth]{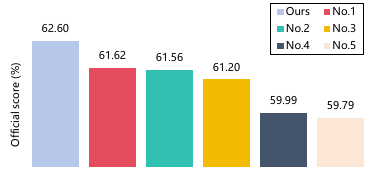}
    \vspace{-0.2cm}
    \caption{Comparison among our system and other  SOTA models on the DCASE 2025 ASD evaluation dataset.}
    \label{fig: result}
    \vspace{-0.2cm}
\end{figure}

Finally, we compare our system against other leading submissions in the DCASE 2025 ASD Challenge. As shown in Fig. \ref{fig: result}, our proposed approach achieves a new SOTA result on the official evaluation dataset\cite{WangMYPS2025,SaengthongSCITOK2025,YangNBU2025,FujimuraNU2025,JiangTHUEE2025}. In addition to its high accuracy, our system is also highly parameter-efficient. Our proposed approach follows the overall framework of our top-ranking system (No.1\cite{WangMYPS2025}), containing only 87M parameters. For comparison, the systems of No.2\cite{SaengthongSCITOK2025}, No.4\cite{FujimuraNU2025}, and No.5\cite{JiangTHUEE2025} are significantly larger, with 569M, 2.38B, and 7B parameters, respectively.

\begin{table}[t]
	\renewcommand\arraystretch{1.25}
	\newcolumntype{L}[1]{>{\raggedright\arraybackslash}p{#1}}
	\newcolumntype{C}[1]{>{\centering\arraybackslash}p{#1}}
	\newcolumntype{R}[1]{>{\raggedleft\arraybackslash}p{#1}}
	\centering
        \footnotesize
	\caption{Ablation study on pseudo labels and model adaption based on the DAP-full scheme.}
	\vspace{-0.3cm}
	\label{tab: ablation}\medskip
	\resizebox{8.5 cm}{!}{\begin{tabular}{c c c c c}
			\toprule[1 pt]
			Pseudo-labels & MA & Dev set & Eval set & All\\
            \midrule
             -&-  & 60.41{\scriptsize$\pm$0.96} 
             & 58.23{\scriptsize$\pm$0.35} 
             & 59.22{\scriptsize$\pm$0.35} \\
             $\checkmark$ &- & \textbf{62.05}{\scriptsize$\pm$0.29} 
             & 60.28{\scriptsize$\pm$0.43} 
             & 61.09{\scriptsize$\pm$0.33} \\
             -& $\checkmark$ & 61.57{\scriptsize$\pm$0.54} 
             & 61.02{\scriptsize$\pm$0.22} 
             & 61.28{\scriptsize$\pm$0.32} \\
            $\checkmark$ & $\checkmark$ & 62.03{\scriptsize$\pm$0.06} 
            & \textbf{62.60}{\scriptsize$\pm$0.33} 
            & \textbf{62.33}{\scriptsize$\pm$0.16} \\
			\bottomrule[1 pt]
	\end{tabular}}
	\vspace{-0.2 cm}
\end{table}

\section{conclusion}
\label{sec:conclusion}

In this paper, we propose a novel ASD approach that leverages attribute-aware representations derived from domain-adaptive, self-supervised pre-training. These powerful representations are then used to generate pseudo-attributes for machines lacking attribute information by employing agglomerative hierarchical clustering with Ward linkage. Furthermore, we perform model adaptation by fine-tuning the domain-adaptive pre-trained model with both pseudo and ground-truth attributes to specialize it for the ASD task. As a result, the proposed method achieves a new SOTA performance on the DCASE 2025 ASD evaluation dataset, outperforming our previous top-ranking system in the challenge.

\vfill\pagebreak

\bibliographystyle{IEEEbib}
\bibliography{strings,refs}

\end{document}